\documentstyle[preprint,aps,prd,eqsecnum]{revtex}
\begin{document}

\preprint{\parbox{40mm}{\large\begin{center}
                        WU~B~95-06 \\[-2mm]
                        NIKHEF~95-006 \\[-2mm]
                        hep-ph/9503418 \\ \end{center}}}

\draft
\title{A PERTURBATIVE APPROACH TO $B$ DECAYS INTO
TWO $\pi$ MESONS}
\author{M.~Dahm, R.~Jakob\footnote{supported by the
Deutsche Forschungsgemeinschaft\\
present address: NIKHEF-K, P.~O.~Box 41882, NL-1009 DB Amsterdam, The
Netherlands}, P.~Kroll
\footnote{e-mail: kroll@wpts0.physik.uni-wuppertal.de}}
\address{Fachbereich Physik, Universit\"at Wuppertal,\\
D-42097 Wuppertal, Germany}
\date{\today}
\maketitle
\begin{abstract}
The modified perturbative approach in which transverse degrees of freedom
as well as Sudakov suppressions are taken into account, is applied to
$B$ decays into two $\pi$ mesons. The influence of various model parameters
(CKM matrix elements, $B$ decay constant, mesonic wave functions) on the
results as well as short distance corrections to the weak Hamiltonian are
discussed in some detail. The perturbative contributions to the $B$ decays
yield branching ratios of the order of $10^{-7}\;-\;10^{-6}$ which values
are well below the upper limit for the $\bar{B}^0\to\pi^+\pi^-$ branching
ratio as measured by CLEO.
\end{abstract}
\newpage

\section{introduction}
\label{sec:intro}
Recently the rare decay modes of the $B$ meson into light mesons, e.g.,
$\bar{B}^0\to\pi^+\pi^-$, attracted much attention of theoreticians although
such decays have barely been observed (see, however, ref.\cite{cleo:94}).
The reason for this interest lies in the expectation that, owing to the large
momentum release, such processes are amenable to a perturbative treatment.
If that is the case the rates for such $B$ decays can be calculated in a
similar fashion as for instance the large momentum transfer behavior
of the pion's electromagnetic form factor. The perturbative approach would
form an attractive alternative to phenomenological models based on soft
physics (e.g., overlaps of wave functions and/or vector meson dominance
\cite{bsw:85}). A reliable estimate of the rates $\Gamma$ for exclusive
non-leptonic $B$ decays is also of enormous importance in the investigation
of CP violation. For the decays of charged $B$ mesons, for instance, one
studies asymmetries between the rates of CP conjugated processes.
A non-zero rate asymmetry, signaling CP violation, requires phase differences
in the interfering amplitudes which may be generated by penguin diagrams.

The first perturbative calculation of exclusive $B$ decays has been
carried out by Szczepaniak, Henley and Brodsky \cite{szc:90}. Other
authors \cite{sim:91,bur:91,car:93,fle:93,hua:94} have repeated that
analysis using similar methods. The results for the decay
widths obtained by the various authors agree with each other to a certain
extent. However, there is a technical difficulty which has not been solved
in a satisfactory way by these authors. As a consequence of the collinear
approximation used in these analyses, i.e.~of the neglect of the transverse
degrees of freedom, a singularity appears in the end-point regions where one
of the partons carries most of the momentum of its parent hadron.
It is presumed that a Sudakov factor damps the end-point regions sufficiently
and a finite, stable result remains. This conjecture is taken as a
justification
for cutting off the end-point regions, a procedure which leads to an unwelcome
strong dependence of the results on the position of the cut-off.

Botts and Sterman \cite{bot:89} have recently calculated a Sudakov factor,
comprising gluonic radiative corrections, for exclusive reactions in
next-to-leading log approximation using resummation and renormalization group
techniques. This approach necessitates the explicit consideration of the
transverse degrees of freedom which, as we mentioned above, are conventionally
neglected. The Botts-Sterman approach which may be termed the modified
perturbative approach, has already been applied to several exclusive reactions,
among them is the pion's and the nucleon's electromagnetic form factor
\cite{lis:92,li:93,jak:93,bol:95,jak:94}. It has turned out that the modified
perturbative approach allows a reliable, theoretically self-consistent
calculation of the perturbative contributions to electromagnetic form factors.
Previous objections \cite{isg:89} against the applicability of the standard
perturbative approach do not apply to the modified approach.

The aim of the present paper is to perform a detailed analysis of $B$ decays
into two $\pi$ mesons using the modified perturbative approach. As it will
turn out, the above-mentioned singularity disappears, hence there is no need
for cutting off the range of integration. A second advantage of the modified
approach is that it allows both a simple parameter-free treatment
of the strong coupling constant and a choice of an appropriate
renormalization scale which keeps under control contributions from higher order
perturbation theory. Owing to these two advantages the modified perturbative
approach provides a reliable estimate of the perturbative contributions to $B$
decays into two $\pi$ mesons. This will be elucidated in some detail in
the body of the text. Comparison with future experimental results will reveal
whether the processes under investigation are indeed dominated by perturbative
contributions or whether they are still under control of soft physics. It
should
be emphasized that of all experimentally accessible exclusive decays into
hadronic final states the process $B\to \pi\pi$ has the largest momentum
release and can therefore be considered as a hard processes best. If here the
perturbative approach fails in comparison with experiment its application
to other exclusive non-leptonic decay processes seems dubious.

In Sec.~2 we discuss the calculation of the decay rate for the process
$\bar{B}^0\to\pi^+\pi^-$ within the modified perturbative approach. The
numerical results and a discussion of a number of approximations made in the
calculation are presented in Sec.~3. Short distance corrections will be
discussed in Sec.~4 as well as a calculation of other $B\to \pi\pi$ decay
modes. The weak $B\to\pi$ transition form factors can be calculated in a
similar fashion as the $B\to\pi\pi$ decay rates. Results for transition form
factors are presented in Sect.~5. The paper terminates with a few concluding
remarks (Sec.~6). In an appendix some details about the Sudakov factor are
presented.

\section{the modified perturbative approach to $B$ decays}
We start with the process, $\bar{B}^0\to\pi^+\pi^-$, and consider the
unperturbed weak Hamiltonian
\begin{equation}
\label{ham2}
{\cal H}_W = 4\frac{G_F}{\sqrt {2}} v_u (\bar u_{\alpha}\gamma_{\mu}L
                   b_{\alpha}) (\bar d_{\beta}\gamma^{\mu}L u_{\beta})
\end{equation}
where $\alpha$ and $\beta$ are color labels. $L$ is the left-handed
projection operator. $G_F$ is the usual Fermi constant and
$v_u = V_{ud}^\ast V_{ub}^{ }$ where the $V_{ij}$ denote
the CKM matrix elements.
Because of the large momentum release in the decay of the $B$ meson
into two light mesons it seems possible to apply perturbative QCD. For the
same reason one may also neglect long-range final state interactions.
On these premises the decay amplitude
${\cal M}=\langle \pi^+\pi^- | {\cal H}_W |\bar{B}^0 \rangle$ can be calculated
along the same lines as for instance the pion's electromagnetic form factor.
With the help of the factorization formula of perturbative QCD for exclusive
reactions the decay amplitude is written as a convolution of meson wave
functions and a hard scattering amplitude $T_H$ to be calculated from the
effective Hamiltonian (\ref{ham2}) and, to leading order in the strong
coupling constant $\alpha_S$, the exchange of a gluon between the spectator
quark and one of the other quarks (see Figs.~1 and 2). In principle there
are contributions from higher Fock states too where additional quarks and/or
gluons appear in the mesons. Such contributions will be ignored by us since
they are suppressed by powers of $\alpha_S/M_B^2$ ($M_B$ denotes the mass of
the $B$ meson). In order to calculate the decay amplitude we utilize the
modified perturbative approach. The crucial element for incorporating the
Sudakov corrections in the calculation of that amplitude within the modified
perturbative approach is the explicit dependence on the transverse
degrees of freedoms in the convolution of wave functions and hard scattering
amplitude. This convolution formula can formally be derived by using the
methods described in detail by Botts and Sterman \cite{bot:89}. Adapting
Li and Yu's result for the semi-leptonic $B\to \pi$ transition matrix
element \cite{li:94} to our case of the $B\to\pi\pi$ decay amplitude,
the convolution formula can be written in the form
\begin{eqnarray}
{\cal M} &=& \frac{G_F}{\sqrt{2}} v_u
\int [{\rm d}x]\left[\frac{{\rm d}^2{\bf b}}{4\pi}\right]
\hat \Psi_{\pi_1}^\ast (x_1,{\bf b}_1) \,
\hat \Psi_{\pi_2}^\ast (x_2,{\bf b}_2) \,
\hat T_H(\{x\},\{{\bf b}\},M_B)
\hat \Psi_B (x,-{\bf b})\nonumber \\
& & \qquad\qquad\times
\exp\left[-S(\{x\},\{b\},M_B)\right].
\label{eq:ampl}
\end{eqnarray}
$x\;(x_i,\,i=1,2)$ denotes the usual fraction of the $p_B^+\;(p_1^+,\,p_2^-)$
component of the $B(\pi_1=\pi^+,\,\pi_2=\pi^-)$ meson's momentum the quark
carries. The antiquarks carry the fractions $1-x$ and $1-x_i$ respectively.
${\bf b}\;({\bf b}_i)$ is the quark-antiquark transverse separation
in the $B\;(\pi_i)$ meson. $[{\rm d}x]$ is short for the product
${\rm d}x{\rm d}x_1{\rm d}x_2$ and $\{x\}$ stands for the set of variables
$x,x_1,x_2$; analogous definitions are used for the
${\bf b}$ variables. Last $\hat \Psi_h (h=B,\pi_i)$ is the
Fourier transform of the momentum space wave function $\Psi_h$:
\begin{equation}
\hat\Psi_h(x_h,{\bf b}_h)= \left(\frac{1}{2\pi}\right)^2
\int {\rm d}^2k_{\perp h} \Psi_h(x_h,{\bf k}_{\perp h}) \,
e^{-i{\bf k}_{\perp h} \cdot {\bf b}_h}
\end{equation}
where the Fourier transform variable ${\bf k}_{\perp h}$ is the quark's
transverse momentum defined with respect to the momentum of the meson $h$.
Strictly speaking the meson wave function represents only the soft part, i.~e.
the full wave function with the perturbative tail removed from it.
The full meson state is described by the product of the (scalar) wave
function $\Psi_h$, a color and a flavor function of obvious form and of the
spin wave function which is written in a covariant fashion
\begin{equation}
\label{cov}
\Gamma_h=\frac{1}{\sqrt{2}}\left(p\hspace{-2.25mm}/_h +M_h\right)
\left(1+\frac{g_h(x_h)}{M_h} \, K\hspace{-3.2mm}/_h \right)
\, \gamma_5.
\end{equation}
$K_h^\mu$ is the relative momentum of quark and antiquark forming the
meson $h$ and $M_h$ is the meson's mass. Note that the relative momentum is
only determined up to a multiple of the hadron's momentum $p_h^\mu$. The term
$\sim K\hspace{-3.2mm}/_h$ in (\ref{cov}) takes into account the fact that
quark and antiquark do not move collinear with their parent hadron; it
represents the first term of an expansion over powers of $K$. The function
$g_h(x_h)$ is controlled by soft physics as is the wave function $\Psi_h$; it
cannot be calculated from first principles at the time being. Therefore, one
has to rely on models if its explicit form is needed. Fortunately, as will
become clear in a moment, we have not to specify $g_h$ to the order we are
working.

In (\ref{eq:ampl}) $\hat T_H$ represents the Fourier transform of the
hard scattering amplitude. To lowest order perturbation theory, it is to be
calculated from the Feynman graphs shown in Fig.~\ref{fig2}. In order to
perform that calculation it is convenient to work in the rest frame of the
$B$ meson where the hadronic momenta in light-cone coordinates read
\begin{equation}
p_B^\mu=(M_B/\sqrt{2},M_B/\sqrt{2},{\bf 0}_\perp),
\qquad
p_1^\mu=(M_B/\sqrt{2},0,{\bf 0}_\perp),
\qquad
p_2^\mu=p_B^\mu-p_1^\mu.
\label{eq:LCmomenta}
\end{equation}
The quark momenta are specified in Fig.~\ref{fig1}. In a possibly crude
approximation, yet well in line with the basic ideas of the parton model, we
chose the relative momenta $K_h^\mu$ in such a way that they have only
transverse components ${\bf k}_{\perp h}$ and, hence, the $K_h^\mu$ are all
orthogonal to the mesonic momenta. The wave function for the $B$ meson is
known to have a pronounced peak at $x=x_0=m_b/M_B$ ($m_b$ being the $b$-quark
mass). It is clear therefore that only regions close to the peak position
contribute to any degree of significance. This implies, to a very good
approximation, that the $b$-quark mass equals $xM_B$. In accord with the
heavy quark effective theory it also implies equal velocities of the $b$-quark
and of the $B$ meson up to corrections of order $k_\perp/M_B$. Since the
r.m.s. transverse momentum is of the order of $300$ to $400\,{\rm MeV}$
these corrections can safely be ignored. The masses of the light quarks
and mesons are neglected in the kinematics.

Given these assumptions and approximations the denominators of the
internal quark and gluon propagators in the graphs Fig.~2a and 2b read
\begin{eqnarray}
\label{prop}
D_b&=&q_b^2-m_b^2= (1-2x+x_1)\;M_B^2-{\bf k}_{\perp 1}^2
\nonumber\\
D_1&=&q_1^2=-(1-x)\;M_B^2-{\bf k}_{\perp}^2
\nonumber\\
q_G^2&=& -(1-x)(1-x_1)\;M_B^2-({\bf k}_{\perp}-{\bf k}_{\perp 1})^2
\end{eqnarray}
where, following previous authors, e.g.,
\cite{szc:90,sim:91,bur:91,car:93,fle:93,hua:94}, we neglect terms
$\sim (1-x)^2$. If one would ignore the $(1-x)$-term in $D_b$ which is
equivalent to the assumption $m_b=M_B$, the denominators of the internal
partons are of the same type as in the case of the pion's electromagnetic form
factor with the $B$-meson mass playing the r\^ole of the momentum transfer.
It has been shown \cite{lis:92,jak:93} that for momentum transfers
of the order of the value of the $B$-meson mass perturbation theory can be
applied self-consistently. In our numerical evaluations of the decay width
we, however, take into account the $(1-x)$-term in $D_b$. It leads to a pole
within the range of integration which is handled in the usual way by
using the prescription $1/(x-i\varepsilon)=P(1/x)+i\pi\delta(x)$.
This pole which corresponds to the situation of the b-quark propagator
going on-shell, is not a pinched singularity and is, therefore, not
associated with long distance propagations of the b quark. Hence, a
perturbative treatment of the pole contribution is justified \cite{col:65}.
For a detailed discussion of this pole we refer to \cite{car:94}.

The hard scattering amplitude $T_H$ can now be easily worked out as traces
of products of the mesonic spin wave functions and spinor expressions
representing the graphs shown in Fig.~\ref{fig2}. From the color structure of
the operator (\ref{ham2}) it is clear that the graphs shown in Figs.~2c and
2d do not contribute to the process $\bar{B}^0\to\pi^+\pi^-$. For graph 2a,
one finds
\begin{equation}
\label{trb}
T_b \sim Tr \left\{\bar{\Gamma}_{\pi_1}\gamma_\mu L
(q_b\hspace{-3.2mm}/+m_b)\gamma_\nu \Gamma_B\gamma^\nu \right\}
     Tr \left\{\bar{\Gamma}_{\pi_2}\gamma^\mu L\right\}
\end{equation}
where, as usual, $\bar{\Gamma}=\gamma_0\Gamma^\dagger\gamma_0$.
Working out the traces we find the simple results
\begin{equation}
T_{b}(\{x\},\{{\bf k}_{\perp}\})=16\sqrt{2}\pi\alpha_s(\mu)
\frac{(2x-x_1)\;M_B^4}{D_bq_G^2}
\left[1+{\cal O}(k_\perp/M_B,k_{\perp 1} /M_B)\right]
\label{tbd}
\end{equation}
and, since we neglect the pion mass, for the graph 2b
\begin{equation}
\frac{|T_{1}|}{|T_{b}|}=
{\cal O}(k_\perp/M_B,k_{\perp 1} /M_B).
\label{eq:t1d}
\end{equation}
$\mu$, the renormalization scale, will be chosen subsequently. A color factor
is not included in $T_b$. The Fourier transform of $T_b$ reads
\begin{eqnarray}
\label{eq:TH-ft}
\hat T_b(\{x\},\{\bf b\})&=&
\frac{4\sqrt{2}}{\pi} \alpha_s(\mu) M_B^4 (2x-x_1)
K_0(\sqrt{(1-x)(1-x_1)}M_Bb) \nonumber\\
&& \times
K_0(\sqrt{2x-x_1-1}M_B|{\bf b}+{\bf b}_1|) \delta({\bf b}_2)
\end{eqnarray}
where $K_0$ is the modified Bessel function of order zero. Note that for
$1+x_1>2x$ the second Bessel function in (\ref{eq:TH-ft}) has
an imaginary argument. In this case we use the relation
$K_0(iz)=-i/2 H_0^{(1)}(z)$ where $H_0^{(1)}$ is a Hankel function. This
treatment of the Fourier transform of $D_b$ is in accord with the
$i\varepsilon$ prescription mentioned earlier.

The last item in (\ref{eq:ampl}) to be specified, is the Sudakov
factor $\exp[-S]$ which comprises those gluonic radiative corrections
not taken into account by the usual QCD evolution. On
the basis of previous work by Collins et al. \cite{col:81}, Botts
and Sterman \cite{bot:89} have calculated the Sudakov factor using resummation
techniques and having recourse to the renormalization group. Adapting
their result to the case at hand, $B \to \pi \pi$, we write
\begin{equation}
\label{sud}
S(x_1,x_2,b_1,b_2,M_B,\mu)= S_{\pi_1}(x_1,b_1,M_B,\mu)
                           +S_{\pi_2}(x_2,b_2,M_B,\mu)
\end{equation}
\begin{equation}
S_{\pi_i}(x_i,b_i,M_B,\mu)= s(x_i,b_i,M_B)+s(1-x_i,b_i,M_B)
          -4/\beta_0 \ln\left(
\frac{\ln(\mu/\Lambda_{QCD})}{\ln(1/(b_i\Lambda_{QCD}))}\right)
\end{equation}
where $\beta_0=11-2n_f/3$; $n_f$ is the number of the active
flavors, which is $4$ in our case. The last term arises from the application
of the renormalization group. The Sudakov function $s$, which includes
leading and next-to-leading logarithms, is given explicitly in the
appendix. Contrary to \cite{li:94} we do not allow for a Sudakov factor for
the $B$ meson. We present the arguments for the ansatz (\ref{sud}) in the
appendix. Although our treatment of the $B$ meson differs from that of
\cite{li:94}, the numerical results are practically the same since the Sudakov
factor for the $B$ meson used in \cite{li:94}, only yields a tiny additional
suppression of the perturbative contribution of the order of a few percent.

Following other authors \cite{lis:92,li:94} we choose as the
renormalization scale the largest mass scale appearing in the process
\begin{equation}
\label{ren}
\mu=\max\left\{\sqrt{(1-x)(1-x_1)}M_B, 1/b,1/b_1\right\}.
\end{equation}
The evaluation of the $\bar{B}^0\to\pi^+\pi^-$ amplitude requires the
integration over the transverse separations. This implies two angle
integrations, say the integration over the directions of ${\bf b}$, simply
providing a factor of $2\pi$, and the integration over the relative
angle $\varphi$ between ${\bf b}$ and ${\bf b}_1$. The latter integration is
non-trivial since the relative angle also appears in the argument of one of
the Bessel functions (see (\ref{eq:TH-ft})) but it can be carried out
analytically by means of Graf's theorem
\begin{eqnarray}
f(y,b,b_1)&\equiv&\frac{1}{2\pi}
\int {\rm d}\varphi K_0(y|{\bf b}\mp{\bf b}_1|)
\nonumber\\
&=&\Theta (b-b_1) K_0(yb)I_0(yb_1) +\Theta (b_1-b) K_0(yb_1)I_0(yb)
\end{eqnarray}
where $\Theta$ is the step function. $I_0(x)=J_0(ix)$, where $J_0$ is
the Bessel function of order zero.
As an inspection of (\ref{eq:TH-ft}) and (\ref{sud}) reveals the $x_2$ and
${\bf b}_2$ integrals factorize. With the help of the $\delta$-function
appearing in (\ref{eq:TH-ft}) the ${\bf b}_2$ integration can directly be
carried out, yielding a factor
\begin{equation}
\frac{1}{4\pi} \int {\rm d}x_2 \hat{\Psi}^\ast_{\pi_2} (x_2,0).
\end{equation}
The value of this integral is fixed by the $\pi^-\to \mu^-\bar{\nu}_\mu$
decay and equals $f_\pi/(2\sqrt{6})$ \cite{lep:83} where $f_\pi$
($=130.7\;{\rm MeV}$) is the pion decay constant. The fact that the
$\pi_2$-part in (\ref{eq:ampl}) factorizes implies the factorization
of non-leptonic decay amplitudes into a product of two current matrix elements
\begin{equation}
\label{fac}
{\cal M} =
\frac{G_F}{\sqrt{2}} v_u
\langle \pi^-(p_2) | J^\mu_W | 0 \rangle
\langle \pi^+(p_1) | J_\mu^W | \bar{B}^0(p_B) \rangle
\end{equation}
with the first matrix element being parameterized in the usual way as
\begin{equation}
\label{pidecay}
\langle \pi^+(p_2) | J^\mu_W | 0 \rangle = f_\pi p_2^\mu.
\end{equation}
Frequently (\ref{fac}) is used as a hypothesis in soft models for decay
amplitudes.

Putting all together we have the following representation of
the $\bar{B}^0\to\pi^+\pi^-$ amplitude
\begin{equation}
\label{eq:finalampl}
{\cal M} =-\frac{\sqrt{3}G_F C_F}{\sqrt{2}}v_u  M_B^4 \Omega_b
\end{equation}
where
\begin{eqnarray}
\label{omb}
\Omega_b&=&\frac{f_\pi}{2\sqrt{3}\pi}
\int {\rm d}x {\rm d}x_1 b{\rm d}b b_1{\rm d}b_1
\hat \Psi_{\pi_1}^\ast (x_1,{\bf b}_1)
\hat \Psi_B (x,-{\bf b})
\exp\left[-S_{\pi_1}(x_1,b_1,M_B,\mu)\right]
\nonumber \\
&& \qquad\times
\alpha_s(\mu)[2x-x_1] K_0(\sqrt{(1-x)(1-x_1)}M_B b) f(\sqrt{2x-x_1-1}M_B,b,b_1)
\end{eqnarray}
The color factor is $\sqrt{3}C_F$ where $C_F(=4/3)$ is the Casimir operator
of the fundamental representation of $SU(3)_c$.
The remaining four dimensional integral has to be carried out
numerically. Before presenting the numerical results (see Sec.~3) two
remarks are in order:\\
i) Owing to the behavior of the $K_0$ function the hard scattering
amplitude has a singularity of the type $\ln(1-x_1)$ for
$x_1\to 1$, $x$ fixed. Since the pion wave function provides a factor
 $1-x_1$ in that limit (see Sec.~3) the integral in (\ref{omb}) is
regular. This result appears as a consequence of the fact that the transverse
degrees of freedom are explicitly considered in the modified perturbative
approach. In the standard approach
\cite{szc:90,sim:91,bur:91,car:93,fle:93,hua:94}, on the other hand,
the neglect of the transverse momenta (and the assumption $m_b=M_B$)
leads to a $1/(1-x_1)^2$ behaviour of the hard scattering amplitude (see
(\ref{prop})). Hence the decay amplitude ${\cal M}$ is logarithmically
singular. That property of $\cal M$ forced previous authors to cut off the
$x_1$-integral at $x_1=1-\varepsilon$ (with $\varepsilon$ being chosen to lie
in the range 0.05-0.1) under the argument that the Sudakov factor will
suppress the end-point region. It should be clear from our discussion that
the singularity of the amplitude is avoided when the transverse degrees of
freedom are taken into account. The Sudakov factor provides only further
suppression of the end-point regions. In contrast to
\cite{szc:90,sim:91,bur:91,car:93,fle:93,hua:94} our results are therefore
insensitive to the end-point regions. Moreover, they do not depend on a
cut-off parameter.\\
ii) For $b(b_1)\Lambda_{QCD}\to 1$, $\alpha_s(\mu)$ is singular. However, as
an inspection of (\ref{eq:sudfct}) reveals, the Sudakov factor tends to zero
faster than any power of $\ln(1/b(b_1)\Lambda_{QCD})$ compensating the
$\alpha_s$ singularity and rendering the integral in (\ref{omb}) finite. This
is an attractive feature the modified perturbative approach possesses; it
allows to choose the renormalization scale in such a way that large logs from
higher order perturbation theory are avoided and a finite result for the
amplitude ${\cal M}$ is nevertheless obtained without introducing
an external parameter (such as e.g. a gluon mass) to regularize
$\alpha_s$. We will take up this point again.

\section{numerical evaluation of the $\bar{B}^0\to\pi^+\pi^-$ decay width}
The first step in the evaluation of that decay width is to specify the
wave functions. The pion wave function is rather well-known since it is
constrained by the decay processes
$\pi^-\to \mu^-\bar{\nu}_\mu$ and $\pi^0\to \gamma\gamma$ \cite{lep:83}.
In \cite{jak:94} a parameterization of that wave function is given
which respects these two constraints.
The soft transverse configuration space wave function is written as
\begin{equation}
\label{eq:wvfct-ansatz}
\hat\Psi_{\pi_i} (x_i,{\bf b}_i) = \frac{f_\pi}{2 \sqrt{6}}
\,\Phi_{\pi_i}(x_i) \,\hat\Sigma_{\pi_i}(\sqrt{x_i(1-x_i)}\,{\bf b}_i).
\end{equation}
For the distribution amplitude $\Phi$ we use
\begin{equation}
\label{eq:DA-AS}
\Phi^{AS}_{\pi_i}(x_i)=6\, x_i(1-x_i)
\end{equation}
and, as an alternative, we use a distribution amplitude originally
proposed by Chernyak and Zhitnitsky \cite{Che:82}
\begin{equation}
\label{eq:DA-CZ}
\Phi_{\pi_i}^{CZ}(x_i)=30\, x_i(1-x_i) \, (2x_i-1)^2 .
\end{equation}
The $b$-dependent part $\hat\Sigma$ is assumed to be a simple Gaussian
\begin{equation}
\label{eq:Sigma-ft}
\hat\Sigma_{\pi_i}(\sqrt{x_i(1-x_i)}\,{\bf b}_i)=
4\pi \; \exp \left( -x_i(1-x_i)\,b_i^2/4 a^2_\pi \right).
\end{equation}
More complicated forms than (\ref{eq:Sigma-ft}) (e.~g., a two-humped shape of
the momentum space wave function) are proposed in \cite{Zhi:93} on the basis
of dispersion relations and duality. At large transverse momentum, however,
the soft momentum space wave function should behave like a Gaussian
\cite{Zhi:93}. The examination of a number of examples corroborates our
expectation that forms of $\hat\Sigma_{\pi_i}$ other than (\ref{eq:Sigma-ft})
will not change the results and the conclusions presented in our
paper markedly. The parameter $a_\pi$ appearing in the Gaussian has a value
of $2.02\,{\rm GeV}^{-1}$ for both the distribution amplitudes.

In \cite{jak:94} it is also shown that the asymptotic (AS) distribution
amplitude combined with the Gaussian (\ref{eq:Sigma-ft}) leads to a reasonable
description of the $\pi\to\gamma$ transition form factor within the modified
perturbative approach while the CZ distribution amplitude seems to be in
conflict with the data. We nevertheless retain that distribution amplitude
since we cannot exclude it definitely and in order to examine the influence
of the form of the distribution amplitude on the final results.

For the $B$ meson we take the Bauer-Stech-Wirbel wave function \cite{bsw:85}
which has been proven to be useful in weak decays. Writing that wave function
in a fashion similar to (\ref{eq:wvfct-ansatz}), we have
\begin{equation}
\label{sigb}
\hat\Sigma_B=
4\pi \; \exp \left( -b^2/4 a^2_B \right)
\end{equation}
and
\begin{equation}
\label{eq:DA-B}
\Phi_B(x)=A\, x(1-x)\,\exp[-a_B^2M_B^2(x-x_0)^2].
\end{equation}
The distribution amplitude $\Phi_B$ exhibits a pronounced peak, its position
is approximately at $x_0=m_b/M_B=0.93$ for $m_b=4.9\,{\rm GeV}$.
This property of the $B$-meson wave function parallels the theoretical
expected and experimentally confirmed behaviour of heavy meson fragmentation
functions. The two parameters appearing in the $B$-meson wave function, namely
$A$ and $a_B$, are fixed by normalizing the wave function to unity - the
neglect of higher Fock states seems to be a reasonable assumption for heavy
hadrons - and by taking a value of $180\,{\rm MeV}$ for the $B$-meson decay
constant $f_B$. This value has been found in a recent lattice gauge theory
analyses \cite{ale:94}. Under these conditions $A$ and $a_B$ have the values
63.05 and $1.491\,{\rm GeV}^{-1}$ respectively. One may be tempted to use a
non-factorizing form for $\Sigma_B$ similarly to that of $\Sigma_\pi$ in
(\ref{eq:Sigma-ft}). However, we favor the function (\ref{sigb}) since the
non-factorizing form has theoretical deficiencies in the formal limit
$M_B \to\infty$ and is, therefore, in conflict with the heavy quark effective
theory as has been shown in \cite{koe:93}.

Using a value of 0.975 for the CKM matrix element $V_{ud}$ and $200\,{\rm MeV}$
for $\Lambda_{QCD}$, we find the following numerical results for the decay
width of the process $\bar{B}^0\to\pi^-\pi^+$ when either the AS distribution
amlitude (\ref{eq:DA-AS}) or the CZ one (\ref{eq:DA-CZ}) is employed:
\begin{eqnarray}
\label{gamma}
\Gamma(\bar{B}^0\to\pi^+\pi^-)=\frac{1}{16\pi M_B} |{\cal M}|^2
                        =\left (\frac{V_{ub}}{0.005}\right)^2
                       \;\times 10^{-10}{\rm eV}
                  \left \{\begin{array} {c@{\quad}c}
                          1.04 & (AS)\\
                          2.58 & (CZ)
                   \end{array} \right.
\end{eqnarray}
Since the CKM matrix element $V_{ub}$ is still poorly known we tentatively take
for it the value 0.005 and show $V_{ub}$ explicitly in (\ref{gamma}).
We repeat that we favor the result obtained with the asymptotic distribution
amplitude and believe that the CZ distribution amplitude perhaps leads to an
overestimate of the perturbative contribution to the $\bar{B}^0\to\pi^+\pi^-$
decay. The predictions for the decay width are also subject to
errors due to uncertainties in the model parameters and assumptions leaving
aside the difference between the results for the two possibilities for the
pion's distribution amplitude. The influence of small modifications in the wave
functions like the use of the BHL factor $\exp[-a_i^2 m_q^2/(x_i(1-x_i))]$
in the distribution amplitudes (where $m_q$ is a constituent mass)
\cite{lep:83}
or changes in the $b$ dependence is rather mild, the decay width is typically
altered by $ 5 - 10 \%$. The largest uncertainty in our results arises
from the error of $50\,{\rm MeV}$ for $f_B$ quoted in \cite{ale:94}.
Although this error is very large and, moreover, the decay width depends
quadratically on $f_B$ , the resulting error for the width only amounts
to about $25\%$. This happens because of the wave function normalisation:
A change in $f_B$ is accompanied by a corresponding change in $a_B$, both
changes compensate each other to a large extent. Combining all uncertainties
in the model parameters we estimate the error of the prediction to amount to
about $35\%$. This errors applies to both the results presented in
(\ref{gamma}).

The results (\ref{gamma}) are comparable in magnitude, although in general
smaller, with previous estimates of the perturbative contribution
\cite{szc:90,sim:91,bur:91,car:93,fle:93,hua:94}
with two exceptions where substantially larger values are quoted. Simma and
Wyler \cite{sim:91} obtain their large value by adjusting the perturbative
contribution to the semi-leptonic decay $B\to De\bar\nu$ to the data and
applying the normalization factor obtained by that procedure also to
non-leptonic $B$ decays. Carlson and Milana \cite{car:93,car:94} use the
so-called peaking approximation advocated for in \cite{koe:93}. In that
approximation the $B$-meson distribution amplitude is assumed to be
$\sim\delta(x-x_0)$ which is with respect to the shape of (\ref{eq:DA-B}).
While the peaking approximation allows one to discuss the qualitative features
of the results in a rather simple fashion it numerically is not very reliable
in some cases. Using it in the evaluation of the decay width within our
approach, we find a result which is larger by a factor of 2.7 (for the AS wave)
function than the value quoted in (\ref{gamma}).

We emphasize that in the modified perturbative approach in which the
transverse degrees of freedom and Sudakov corrections are taken
into account, the soft end-point regions are strongly suppressed. Therefore,
there is no need for an cut-off parameter in the $x_1$ integration as it is
the case in the previous perturbative calculations of the $\bar{B}^0$ decay
width \cite{szc:90,sim:91,bur:91,car:93,fle:93,hua:94}. The extreme
sensitivity of the results to the region near $x_1=1$ found by these authors
has completely disappeared in our approach. There is second advantage in
our approach: we can make use of the standard one-loop formula for the
strong coupling constant. Moreover, we can choose the renormalization scale
as to avoid large logs from higher order perturbation theory (see (\ref{ren})).
Still the integral appearing in (\ref{omb}) is regular since the Sudakov factor
compensates the $\alpha_S$ singularity. Actually the suppression of the
end-point regions is so strong that the bulk of the perturbative contribution
is accumulated in regions where $\alpha_S$ is small. In order to demonstrate
that we cut off in (\ref{omb}) regions where $\alpha_S$ exceeds a given value
$\alpha_c$ and plot $\Omega_b$ as a function of $\alpha_c$. As can be seen from
Fig.~3 $50\%$ of the result is already accumulated
in regions where $\alpha_S$ is smaller than about 0.5. The result
saturates for $\alpha_S \simeq 0.8$, there are practically no contributions
from regions where $\alpha_S$ is larger than that value. Therefore, we
consider the perturbative contribution to the decay amplitude as theoretically
self-consistent. Note that the regions of large values of $\alpha_S$ correspond
to large quark-antiquark separations in the mesons. The use of the one-loop
formula for $\alpha_S$ (as well as the parameterization of the parton
propagators) in the end-point regions may be questioned. However, the
saturation property of the perturbative result tells us that
$\alpha_S$ can be frozen in at a value $\simeq 0.8$ or regularized by,
say, the introduction of a gluon mass without changing the final results.
In order to avoid the introduction of such external parameters we keep
the standard $\alpha_S$ parameterization, being conscious of the fact that
the actual parameterization is of no account in the soft end-point regions.
%

\section{short distance corrections}
It has been shown (see, for instance, \cite{bur:92,ali:94} and references
therein) that leading order short distance corrections lead to an effective
Hamiltonian at the scale $\mu$
\begin{equation}
\label{ham6}
{\cal H}_W^{eff} = 4\frac{G_F}{\sqrt {2}} v_u \sum_{i=1,2} C_i(\mu) {\cal O}_i
                 + 4\frac{G_F}{\sqrt {2}} v_t \sum_{i=3}^{6}
                   C_i(\mu) {\cal O}_i,
\end{equation}
replacing the weak Hamiltonian (\ref{ham2})
($v_t=V^\ast_{td}V_{tb}^{ }=-v_u - v_c$). The operators ${\cal O}_i$ are
given as follows
\begin{eqnarray}
\label{oper}
{\cal O}_1&=&(\bar d_{\alpha}\gamma_{\mu}L b_{\alpha})
                    (\bar u_{\beta}\gamma^{\mu}L u_{\beta})
                      \quad\quad\hspace{5mm}
{\cal O}_2 = (\bar d_{\beta}\gamma_{\mu}L b_{\alpha})
                    (\bar u_{\alpha}\gamma^{\mu}L u_{\beta}) \nonumber\\
{\cal O}_3&=&(\bar d_{\alpha}\gamma_{\mu}L b_{\alpha})
                   \sum_q (\bar q_{\beta}\gamma^{\mu}L q_{\beta}) \quad\quad
{\cal O}_4 = (\bar d_{\beta}\gamma_{\mu}L b_{\alpha})
                   \sum_q (\bar q_{\alpha}\gamma^{\mu}L q_{\beta}) \nonumber\\
{\cal O}_5&=&(\bar d_{\alpha}\gamma_{\mu}L b_{\alpha})
                   \sum_q (\bar q_{\beta}\gamma^{\mu}R q_{\beta}) \quad\quad
{\cal O}_6 = (\bar d_{\beta}\gamma_{\mu}L b_{\alpha})
                   \sum_q (\bar q_{\alpha}\gamma^{\mu}R q_{\beta})
\end{eqnarray}
where $R$ denotes the right-handed projection operator and $q$ is running
over the quark flavors being active at the scale $\mu$. The Wilson
coefficients at the scale $\mu=M_W$ are $C_k(M_W)=\delta_{k2}$. The
renormalization group evolution from the scale $M_W$ to $\mu=m_b$, relevant
to the $B$ decays, leads
to the following values of the Wilson coefficients at the latter scale
\cite{ali:94}:
\begin{eqnarray}
\label{wil}
C_1=-0.229, \quad &\quad C_2=1.097, \quad &\quad C_3=\;\;\;0.021,\nonumber\\
C_4=-0.039, \quad &\quad C_5=0.007, \quad &\quad C_6=-0.029.
\end{eqnarray}
At the scale $\mu=M_W$ (\ref{ham6}) reduces to the Hamiltonian (\ref{ham2});
after Fierz reordering the operator ${\cal O}_2$ is the one appearing in
(\ref{ham2}).

Taking into account the Hamiltonian (\ref{ham6}) in the calculation of
$B$ decays into two pions, we have to consider many contributions. There
is, on the one side, the class of direct contributions where the $d$ quark
together with the spectator quark forms one of the pions. A second class
contains the exchanged contributions where the $d$ quark forms a pion
together with one of the other light quarks. Due to Fierz reordering,
however, the exchanged contributions can be brought into the form of the
contributions from the first class, with eventually different color
factors and/or Wilson coefficients. Consequently, we have only to work
out three Dirac traces and hence three integrals corresponding to the
graphs shown in Figs.~2a, 2c and 2d. As we mentioned in Sect.~2 the graph 2b
leads to a vanishing contribution for zero pion mass. The relevant trace and
the integral for graph 2a are given in (\ref{trb}) and (\ref{omb}).

Making use of the same set of approximations as in Sect.~2,
we find for the contributions from the graphs 2c and 2d:
\begin{eqnarray}
\label{om2}
\Omega_2&=&\frac{\sqrt{2}}{4\pi^2}
\int [{\rm d}x] b{\rm d}b b_2{\rm d}b_2
\hat \Psi_{\pi_1}^\ast (x_1,-{\bf b})\hat \Psi_{\pi_2}^\ast (x_2,{\bf b}_2)
\hat \Psi_B (x,-{\bf b})\nonumber\\
&&\qquad\times\exp\left[-S_{\pi_1}(x_1,b,M_B,\mu)
                        -S_{\pi_2}(x_2,b_2,M_B,\mu)\right]
                 \alpha_s(\mu)[x_2+x-1]
\nonumber \\
&& \qquad\times
K_0(\sqrt{(1-x)(1-x_1)-(x-x_1)x_2}M_B b)
                                        f(\sqrt{(1-x)(1-x_1)}M_B,b,b_2)
\end{eqnarray}
and
\begin{eqnarray}
\label{om3}
\Omega_3&=&\frac{\sqrt{2}}{4\pi^2}
\int [{\rm d}x] b{\rm d}b b_2{\rm d}b_2
\hat \Psi_{\pi_1}^\ast (x_1,-{\bf b}) \hat \Psi_{\pi_2}^\ast (x_2,{\bf b}_2)
\hat \Psi_B (x,-{\bf b})
\nonumber \\
&& \qquad\times
\exp\left[-S_{\pi_1}(x_1,b,M_B,\mu)
          -S_{\pi_2}(x_2,b_2,M_B,\mu)\right]
          \alpha_s(\mu)[x_1+x_2-2x]
\nonumber \\
&& \qquad\times
        K_0(\sqrt{1-x_1x_2-x(2-x_1-x_2)}M_B b)
                              f(\sqrt{(1-x)(1-x_1)}M_B,b,b_2).
\end{eqnarray}
In the renormalization scale expression (\ref{ren}) $b_1$ is to be replaced by
$b_2$. Obviously, the contributions from the penguin operators ${\cal O}_3$
and ${\cal O}_4$ lead to the same integrals, (\ref{omb}), (\ref{om2}) and
(\ref{om3}) as the operators ${\cal O}_1$ and ${\cal O}_2$ respectively. The
contributions from the operators ${\cal O}_5$ and
${\cal O}_6$, involving right-handed quarks, are either zero or lead again
to the integrals (\ref{omb}), (\ref{om2}) and (\ref{om3}).

There is a third class of contributions, namely those from annihilation
topologies (see Fig.~4). In principle such topologies are generated by
all six operators ${\cal O}_i$  and in each of these topologies a gluon can be
exchanged between the quark line connecting $\pi_1$ and $\pi_2$,
and any of the other quarks, i.~e., there are again four graphs contributing
similar to those shown in Fig.~2. The contributions from the factorizing
graphs are, however, zero. Only the contributions from the graphs where the
gluon is exchanged between one of the quarks forming the $B$ meson and the
quark connecting $\pi_1$ and $\pi_2$ are non-zero provided the color structure
is appropriate. The two non-zero contributions read
\begin{eqnarray}
\label{om3a}
\Omega_b^a&=&\frac{\sqrt{2}}{4\pi^2}
\int [{\rm d}x] b{\rm d}b b_2{\rm d}b_2
\hat \Psi_{\pi_1}^\ast (x_1,{\bf b}_2) \hat \Psi_{\pi_2}^\ast (x_2,{\bf b}_2)
\hat \Psi_B (x,-{\bf b})
\nonumber \\
&& \qquad\times
\exp\left[-S_{\pi_1}(x_1,b_2,M_B,\mu)
          -S_{\pi_1}(x_2,b_2,M_B,\mu)\right]
          \alpha_s(\mu)[x_1-1]
\nonumber \\
&& \qquad\times
        K_0(\sqrt{(1-x)(1-x_1-x_2)-x+x_1x_2)}M_B b)
                              f(\sqrt{(1-x_1)(1-x_2)}M_B,b,b_2) \nonumber\\
\Omega_1^a&=&\frac{\sqrt{2}}{4\pi^2}
\int [{\rm d}x] b{\rm d}b b_2{\rm d}b_2
\hat \Psi_{\pi_1}^\ast (x_1,{\bf b}_2) \hat \Psi_{\pi_2}^\ast (x_2,{\bf b}_2)
\hat \Psi_B (x,-{\bf b})
\nonumber \\
&& \qquad\times
\exp\left[-S_{\pi_1}(x_1,b_2,M_B,\mu)
          -S_{\pi_2}(x_2,b_2,M_B,\mu)\right]
          \alpha_s(\mu)[x-x_1]
\nonumber \\
&& \qquad\times
        K_0(\sqrt{(x-x_1)(x-x_2)}M_B b)
                              f(\sqrt{(1-x_1)(1-x_2)}M_B,b,b_2).
\end{eqnarray}

Putting all together, we find the following expressions for the decay
amplitudes of the three $\pi\pi$ channels
\begin{eqnarray}
\label{final}
{\cal M}(\bar{B}^0\to\pi^+\pi^-)&=&
                   -\frac{G_FC_F}{\sqrt{6}} v_u M_B^4
              \left\{[C_2-v_t/v_uC_4]3\Omega_b \right.
\nonumber\\
     &&+ [C_1-v_t/v_uC_3][\Omega_b + \Omega_2 +\Omega_3]
     \left.   + [C_2-2v_t/v_uC_4][\Omega^a_b+\Omega^a_1]\right\}
\nonumber\\
{\cal M}(B^-\to\pi^0\pi^-)&=&
                  - \frac{G_FC_F}{2\sqrt{3}} v_u M_B^4
                [C_1+C_2] \left[4\Omega_b +\Omega_2+\Omega_3)\right]
\end{eqnarray}
and
\begin{equation}
\label{finalb}
{\cal M}(\bar{B}^0\to\;\pi^0\;\pi^0)=1/\sqrt{2}{\cal M}(\bar{B}^0\to\pi^+\pi^-)
                                    -{\cal M}(B^-\to\pi^0\pi^-)
\end{equation}
i.~e.~the amplitudes satisfy the familiar isospin relation for $B\to\pi\pi$
decays \cite{gro:90}. The contributions generated by the operators
${\cal O}_5$ and ${\cal O}_6$ are either zero or so small (for annihilation
topologies) that they can safely be neglected.

Numerical evaluation of the amplitudes and the corresponding rates
using the wave functions discussed in Sect.~3, leads to the final results
for the decay widths for the three $B$ decay processes shown in Table 1.
The results shown in the table are evaluated with $V_{ub}=0.005$ and
$v_t=-0.0138+i0.0050$. The uncertainties due to the wave functions and the
$B$-decay parameter amount to about $35\%$, see the discussion in Sect.~3.
As an inspection of Table 1 reveals the QCD corrections are substantially
and, in total, amount to a reduction of the leading term generated by the
Hamiltonian (\ref{ham2}) of $20 - 25 \%$. An exceptional r\^ole plays
the $\bar{B}^0$ decay into two uncharged pions because for that reaction
there is no direct contribution from the Hamiltonian (\ref{ham2}) and the
exchanged contribution is suppressed by color (a factor 1/3 in the amplitude).
The other contributions are therefore relatively strong. The penguin
contributions provide corrections of the order of  $20\%$ to the process
$\bar{B}^0\to\pi^+\pi^-$. Since $v_t$ is complex the CP conjugated $B^0$ decay
rates slightly differ from the $\bar{B}^0$ rates presented in the table
(by $\simeq0.2\%$). We stress that our aim is to estimate, as accurate
as possible, the strength of the perturbative contributions to the $B$
decay rates. For a calculation of such subtle effects as CP violations
a more refined treatment of the penguin contributions, taking into account
${\cal O}(\alpha_S^2)$ corrections, is required \cite{sim:91,fle:93,kra:94}.
Our conclusions about the strength of the perturbative contributions to the
$B$ decay rates will not be altered substantially by a refined treatment of
the penguin contributions. Note that in other processes, in particular if
$b\to s$ transitions are involved, the r\^ole of the penguin graphs may be
more important (see e.~g.~\cite{sim:91}). There are also reactions in which
the annihilation contributions are more important than in the $B\to\pi\pi$
decays, e.~g., in $B^-\to K^0K^-$ where the direct and exchanged
contributions are only generated by the penguin operators.

Our final predictions for the branching ratios are:
\begin{eqnarray}
\label{finr}
BR(\bar{B}^0\to\pi^+\pi^-)&=& \left (\frac{V_{ub}}{0.005}\right)^2 \;
                  \left \{\begin{array} {c@{\quad}c}
                          0.16 & (AS)\\
                          0.44 & (CZ)
                   \end{array} \right \}\times 10^{-6} \pm 35\%, \nonumber\\
BR(\bar{B}^0\to\pi^0\pi^0)&=& \hspace{2.0cm}
                  \left \{\begin{array} {c@{\quad}c}
                          0.69 & (AS)\\
                          0.71 & (CZ)
                   \end{array} \right \}\times 10^{-8},\nonumber\\
BR(\bar{B}^-\to\pi^0\pi^-)&=& \left (\frac{V_{ub}}{0.005}\right)^2 \;
                  \left\{
                           \begin{array} {c@{\quad}c}
                          0.15 & (AS)\\
                          0.36 & (CZ)
                   \end{array} \right\}\times 10^{-6} \pm 35\%.
\end{eqnarray}
We emphasize again that we favor the results obtained with the AS
wave function for reasons explained in \cite{jak:94} yet the other results
obtained with the CZ wave function cannot definitely be excluded for the
time being. The errors quoted in (\ref{finr}) do not contain the uncertainty
in the value of $V_{ub}$ but it is shown explicitly in (\ref{finr}). Since
the penguin operators provide only small corrections also the branching ratio
for the first reaction is (approximately) proportional to $V_{ub}^2$. Since
the rate for the $\pi^0\pi^0$ channel is much smaller than the other rates
we refrain from quoting an error for it.

While the decay of the $\bar{B}^0$ meson into two charmless mesons has been
observed experimentally the results for individual branching ratios still
suffer from large statistical and systematical errors. In fact only an upper
limit of $2.9\times 10^{-5}$ for the process $\bar{B}^0\to\pi^+\pi^-$ has
been quoted by the CLEO collaboration. Our predictions are an order of
magnitude below that limit, tempting us to conclude that the perturbative
contributions to the $B\to \pi\pi$ decays are likely be too small, soft
physics may still dominate these processes. Yet a definite conclusion cannot
be drawn at present.

For comparison we also quote the relative rates for $\bar{B}^0$ decays
into two-pion states with isospin 2 and 0
\begin{equation}
\frac{|\langle \pi\pi, I=2 \mid{\cal H}_W^{eff} \mid \bar{B}^0 \rangle|^2}
     {|\langle \pi\pi, I=0 \mid{\cal H}_W^{eff} \mid \bar{B}^0 \rangle|^2}
                    = \left \{\begin{array} {c@{\quad}c}
                          0.84 & (AS)\\
                          0.71 & (CZ)
                   \end{array}\right.
\end{equation}
We observe that perturbation theory slightly favors $\Delta I =1/2$
transitions over $\Delta I =3/2$ transitions in $B$ decays.

\section{$B$-$\pi$ transition form factors}
Our calculations presented in Sect.~2 and 3 implicitly contain an
evaluation of the $B$-$\pi$ transition form factors at $p_2^2=0$. These
calculations can easily be generalized to other values of $p_2^2$.
The transition form factors are defined as
\begin{equation}
\langle \pi^+(p_1) | J_\mu^W | \bar{B}^0(p_B) \rangle =
  F_{+}\left(p_2^2\right) (p_B+p_1)^\mu
+ F_{-}\left(p_2^2\right) (p_B-p_1)^\mu.
\label{ME-Btopi}
\end{equation}
We write the energy of the $\pi^+$ meson in the $B$ meson rest frame as
\begin{equation}
E_{\pi_1}=\eta M_B/2
\end{equation}
where $0\leq \eta \leq 1$. Then the momentum transfer $p_2^2$ equals
$M_B^2(1-\eta)$. Calculating the two form factors from the graphs 2a
and 2b, we find
\begin{eqnarray}
F_+&=& \frac{1}{2} \frac{C_FM_B^2}{\pi}
\left(\frac{2\sqrt{3}\pi}{f_\pi}\Omega_b-(1-\eta)\Omega_1 \right)
\nonumber\\
F_-&=& \frac{1}{2} \frac{C_FM_B^2}{\pi}
\left(-\frac{2\sqrt{3}\pi}{f_\pi}\Omega_b+(1+\eta)\Omega_1 \right)
\nonumber\\
\end{eqnarray}
where the integral $\Omega_b$ is given in (\ref{omb}) with the replacements:
$M_B$ by $\sqrt{\eta} M_B$ and $[2x-x_1]$ by
$[2x-1+(1-x_1)\eta]$. The contribution from the graph 2b leads to an
integral similar to $\Omega_b$
\begin{eqnarray}
\label{om1}
\Omega_1&=&
\int {\rm d}x {\rm d}x_1 b{\rm d}b b_1{\rm d}b_1
\hat \Psi_{\pi_1}^\ast (x_1,{\bf b}_1)
\hat \Psi_B (x,-{\bf b})
\exp\left[-S_{\pi_1}(x_1,b_1,M_B,\mu)\right]
\nonumber \\
&& \qquad\times
\alpha_s(\mu)[1-x] K_0(\sqrt{(1-x)(1-x_1)\eta} M_B b)
f(\sqrt{(1-x)\eta}M_B,b,b_1).
\end{eqnarray}
The results for the transition form factors obtained with the wave
functions (\ref{eq:wvfct-ansatz}), (\ref{eq:DA-AS}) and (\ref{eq:Sigma-ft})
for the pion and (\ref{sigb}), (\ref{eq:DA-B}) for the $B$ meson, are
shown in Fig.~5. The results can be trusted for values of $\eta$ between
0.2 and 1. For smaller values of $\eta$ the perturbative contribution is
inconsistent in the sense that more than 50\% of the perturbative contribution
is built up in regions where $\alpha_s \geq 0.7$. The perturbative contribution
to the differential decay rate for the semileptonic $B\to\pi$ decay does not
match at $\eta=0.2$ with the soft pion result obtained in \cite{bur:94}
and, integrated from $\eta=0.2$ to $\eta=1.0$ is tiny in comparison with
the current experimental limit \cite{ong:93} on the branching ratio
of $\bar{B}^0\to\pi^+l^-\bar{\nu}_l$.

Li and Yu \cite{li:94} have also calculated these form factors within the
modified perturbative approach. While we agree with their analytical
expressions for the form factors, our numerical results are about an order of
magnitude smaller than theirs. We suspect that Li and Yu normalised the $B$
state wrongly.

Several other predictions for the form factor $F_+$ at $\eta =1$ are to be
found in the literature (BSW overlap model\cite{bsw:85}, QCD sum rules
\cite{bal:91,bel:93}, lattice gauge theory \cite{all:93}). The
predicted values from these non-perturbative approaches range between
0.24 and 0.33. For the form factor $F_+$ at $\eta =1$ the decay width
$\Gamma(B^0\to\pi^+\pi^-)$ can be estimated, assuming factorization
(\ref{fac}) to hold for soft physics. One finds
\begin{equation}
\Gamma(\bar{B}^0\to\pi^+\pi^-)= \frac{G_F^2 |v_u|^2}{32\pi}
                          f_\pi^2 \, M_B^3 \, |F_+(0)|^2.
\end{equation}
and using the quoted values for $F_+(\eta=1)$, the predictions for the
branching ratio range from $0.9\times 10^{-5}$ to $1.8\times 10^{-5}$, i.e.,
they are rather close to the upper limit measured by the CLEO collaboration
\cite{cleo:94} in contrast to the perturbative contribution.

\section{concluding remarks}
We have calculated the rates for $B\to\pi\pi$ decays in the modified
perturbative approach in which the transverse degrees of freedom as well as
Sudakov effects, comprising gluonic radiative corrections, are taken into
account. We believe that the perturbative contributions to these processes
are reliably estimated: The hard scale is provided by $M_B^2$ and the
Sudakov factor suppresses the soft end-point regions strongly so that the
bulk of the perturbative contributions is accumulated in regions where
$\alpha_S$ is sufficiently small. Therefore our estimate of the perturbative
contribution can be considered as theoretically self-consistent. The
difficulties previous authors \cite{szc:90,sim:91,bur:91,car:93,fle:93,hua:94}
had with the singular behaviour of the hard scattering amplitude disappear
in our approach, there is no need for a cut-off and correspondingly no need
for an additional external free parameter. Also the extreme sensitivity of the
perturbative contribution to the (soft) end-point regions disappears
completely. The phenomenological input into our calculation, namely the
mesonic wave functions - which are controlled by long distance physics and
are, therefore, not calculable at present to a sufficient degree of accuracy
- is fairly well constrained. The influence of various parameters and
corrections on our results is discussed in some detail.
The major uncertainties arise from the CKM matrix element $V_{ub}$, the
$B$-meson decay constant $f_B$ and from the pion's distribution amplitude
(although we favor the results obtained with the AS distribution amplitude).
The experimental upper bound for the $\bar{B}^0\to\pi^+\pi^-$ branching ratio
is rather large as compared with the magnitude of the perturbative
contribution. Although definite conclusions cannot be drawn as yet, we have
to be aware that $B\to\pi\pi$ decays are perhaps controlled by soft physics.
Estimates of rate asymmetries based on perturbation theory may be misleading.

In a similar fashion as the $B\to\pi\pi$ decays we can also calculate the
rates for $B$ decays into other light mesons involving $K$, $\rho$ and/or
$K^*$ mesons. Depending on the particular channel considered the r\^ole
of the operators in (\ref{ham6}) other than ${\cal O}_2$, may be more
important than in the $\pi \pi$ channel. Thus, as shown in \cite{sim:91},
the penguin graphs provide substantial corrections to the process
$B^-\to K^-\pi^0$. We expect similarly small perturbative contributions to
the rates of other light meson channels. Of utmost phenomenological interest
are $B$ decays into channels involving $D$ mesons since the branching
ratios for these processes are typically two orders of magnitude larger than
the upper bound for the $\bar{B}^0\to\pi^+\pi^-$ branching ratio.
This enhancement is essentially due to the fact that now the CKM matrix
element $V_{bc}$ is involved instead of the much smaller matrix element
$V_{ub}$. A treatment of such channels within the modified perturbative
approach is justified since, in contrast to the standard approach, the
relevant parton virtualities are sufficiently large, and correspondingly
$\alpha_S$ is small enough, thanks to the transverse degrees of freedom (see
(\ref{prop}) and (\ref{ren})). In the light of our experience with the
$\pi \pi$ channels and backed by explorative studies we do not expect that
the perturbative contributions to the channels involving $D$ mesons are
large enough to account for the experimental values.

\acknowledgments
We thank J.~Bolz for providing us with the corrected form of the Sudakov
function. We acknowledge useful discussions with T.~Mannel.

\appendix
\section*{}
In this appendix we present a few details on the Sudakov factor
used in our calculation. The Sudakov factor comprises those parts of
gluonic radiative corrections which are not taken into account by the usual
QCD evolution. Characteristic of it are double logs produced by overlapping
collinear and soft divergencies (for almost massless quarks). Examples of
one-loop graphs responsible for radiative corrections are shown in
Fig.~\ref{fig6}. In axial gauges the two-particle reducible graphs, like the
examples shown in Fig.~\ref{fig6}a, give rise to double logs, whereas the
non-reducible graphs (Fig.~\ref{fig6}b) only lead to single logs. The double
logs from higher order loops can be resummed using the techniques developed
by Collins et al.~\cite{col:81}. The resummation results in the exponentiated
one-loop corrections.

We analyse the process under consideration, $B\to\pi\pi$, in the
$B$ meson rest frame (see(\ref{eq:LCmomenta})). In this frame the pions
have momenta with either a large $+$ or a large $-$ light-cone component,
while the other components are zero. The same uneven kinematical situation
holds for the quarks the pions are made up. This is exactly the kinematical
situation for which Collins et al.~\cite{col:81} derived the Sudakov factor.
For exclusive processes, like $B\to\pi\pi$, the Sudakov factor or more
precisely the Sudakov function reads \cite{bot:89}
\begin{eqnarray}
s(\xi,b,Q)&= & \frac{8}{3\beta_0}
\left(\hat q \ln\left(\frac{\hat q}{\hat b}\right)
      -\hat q+\hat b\right)
\nonumber\\
&+& \frac{4\beta_1}{3\beta_0^3}
\left[\hat q
\left(\frac{\ln(2\hat q)+1}{\hat q}-\frac{\ln (2\hat b)+1}{\hat b}\right)
+\frac{1}{2}
\left(\ln^2(2\hat q)-\ln^2(2\hat b) \right)
\right]
\nonumber\\
&+& \frac{4}{3\beta_0}\ln\left(\frac{e^{2\gamma-1}}{2}\right)
       \ln\left(\frac{\hat q}{\hat b}\right)
                  +A^{(2)}\frac{4}{\beta_0^2}
\left[
\frac{\hat q-\hat b}{\hat b}-\ln\left(\frac{\hat q}{\hat b}\right)
\right]
\label{eq:sudfct}
\end{eqnarray}
where the definitions
\begin{equation}
\hat q \equiv \ln\left(\xi Q/\sqrt{2}\Lambda_{QCD}\right);
\quad  \quad
\hat b \equiv \ln\left(1/b\Lambda_{QCD}\right)
\end{equation}
and
\begin{equation}
A^{(2)}\equiv\frac{67}{9}-\frac{\pi^2}{3}-\frac{10}{27}n_f
+\frac{2\beta_0}{3}\ln\left(e^\gamma/2\right)
\end{equation}
are used ($\beta_1=102-38n_f/3$; $\gamma$
is the Euler constant). $Q/\sqrt{2}$ is the
large ($+$ or $-$) component of the pion's momentum for an appropriate
choice of the gauge vector. In the case of
$B\to\pi\pi$ (see (\ref{eq:LCmomenta})) $Q$ equals $M_B$ and for the $B$-$\pi$
transition form factor $Q=\eta M_B$. $\xi$ is the relevant
momentum fraction $x_i$ or $(1-x_i)$, $i=1,2$. Note that the
coefficients in the second line of (\ref{eq:sudfct}) differ from previous
publications\cite{bot:89,lis:92,li:93}. In the case of the pion
form factor the effect of these corrections to the Sudakov function
results in a reduction of the perturbative contribution of about 2\% as
compared to previously reported results \cite{jak:93}.

The range of validity of (\ref{eq:sudfct}) for the Sudakov function
is limited to not too small values of the transverse separation
of quark and antiquark in the meson. Whenever $b\leq \sqrt{2}/\xi Q$
(i.e., $\hat b\geq\hat q$), the gluonic corrections are to be
considered as higher-order corrections to the hard scattering amplitude and,
hence, are not contained in the Sudakov factor but are absorbed in $\hat T_H$.
Therefore,
\begin{equation}
s(\xi,b,Q)=0
\qquad {\rm if} \quad
\hat b\geq \hat q
\label{eq:sudzero}
\end{equation}
is assumed. Similarly, the complete Sudakov factor $\exp[-S]$ is set to
unity, whenever it exceeds unity, which is the case in the small $b$ region.
As $b$ increases the Sudakov function increases as well, tending to infinity
with $b\to 1/\Lambda_{QCD}$. Consequently, the Sudakov factor $\exp[-S]$
drops to zero. For $b_1$ larger than $1/\Lambda_{QCD}$, the
true soft region, the Sudakov factor is zero.

For the $B$ meson the situation is completely different from that
of the $\pi$ mesons. Due to the large $b$-quark mass the radiative
corrections only produce soft divergencies but no collinear
ones. Consequently, double logs do not appear and, hence, the Sudakov
function for the $b$-quark is zero. For the light quark there are
still both soft and collinear divergencies. However, because of
the pronounced peak the $B$-meson wave function exhibits, the bulk of
the perturbative contribution is accumulated in regions where the
momentum of the light quark is soft, i.e., has no large component at
all. As has been shown in \cite{col:81} the divergencies do not
overlap in these situations. Thus, also for the light quark the
Sudakov function is zero.\\
In contrast to us, Li and Yu \cite{li:94} also associate a Sudakov function
to the light quark contained in the $B$ meson. Because of the
shape of the $B$-meson wave function only values of $x$ close to $x_o$ are
relevant for which there is only a weak or even no Sudakov suppression.
Thus, wether or not a Sudakov function for the light quark in the $B$ meson is
taken into account is irrelevant in praxis, the numerical values
for the decay rates differ only by 5\%.

\newpage


\clearpage
%
%
\begin{table}
\caption[tabledummy]
{The rates for $B \to\pi\pi$ decays in units of $10^{-10}\,{\rm eV}$ as
predicted from the effective Hamiltonian (\ref{ham6}). Column A: Contributions
from the operator ${\cal O}_2 (C_2=1)$, annihilations neglected. Column B:
Contributions from ${\cal O}_1$ and ${\cal O}_2$, annihilations neglected.
Column C: As B but penguin contributions included. Column D: Full result with
annihilations included.}
\begin{center}
\label{table1}
\begin{tabular}{lccccc}
 &  & A  & B & C & D \\
\hline
           & AS & 1.04 & 1.11 & 0.92 & 0.81 \\
\raisebox{1.5ex}[-1.5ex]{$\bar B^0\to\pi^+\pi^-$}
           & CZ & 2.58 & 2.88 & 2.30 & 2.21 \\
\hline
           & AS & 0.050 & 0.015 & 0.021 & 0.034 \\
\raisebox{1.5ex}[-1.5ex]{$\bar B^0\to\pi^0\pi^0$}
           & CZ & 0.089 & 0.012 & 0.043 & 0.035 \\
\hline
           & AS & 0.86 & 0.65 & 0.65 & 0.65 \\
\raisebox{1.5ex}[-1.5ex]{$B^-\to\pi^0\pi^-$}
           & CZ & 2.03 & 1.53 & 1.53 & 1.53
\end{tabular}
\end{center}
\end{table}

\clearpage
%
%
\begin{figure}
\caption[dummy1]
{The basic graph for the decay of a $B$ meson into two pions. The circle
stands for the effective weak Hamiltonian. The quark momenta are specified.}
\label{fig1}
\end{figure}

\begin{figure}
\caption[dummy2]
{Lowest order Feynman graphs for $B$ decays into two pions. The internal
quark and gluon momenta are indicated.}
\label{fig2}
\end{figure}

\begin{figure}
\caption[dummy3]
{Saturation of the perturbative contribution: $\Omega_b$ vs.~$\alpha_c$
(see text). Solid (dashed) line represents the perturbative contribution
with (without) the Sudakov factor.}
\label{fig3}
\end{figure}

\begin{figure}
\caption[dummy4]
{The annihilation topology.}
\label{fig4}
\end{figure}

\begin{figure}
\caption[dummy5]
{The $B$-$\pi$ transition form factors vs.~$\eta$. The solid (dashed) and
dash-dotted (dotted) lines represent the results for the form factor $F_+$
($F_{-}$) obtained with the AS and CZ wave functions respectively.}
\label{fig5}
\end{figure}

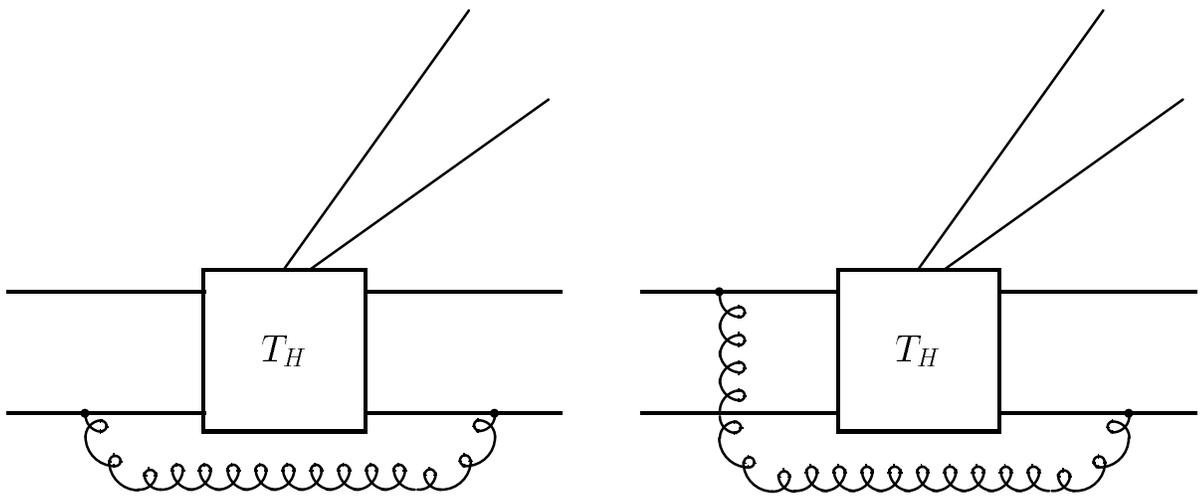
\begin{figure}
\caption[dummy6]
{Radiative corrections.}
\label{fig6}
\end{figure}

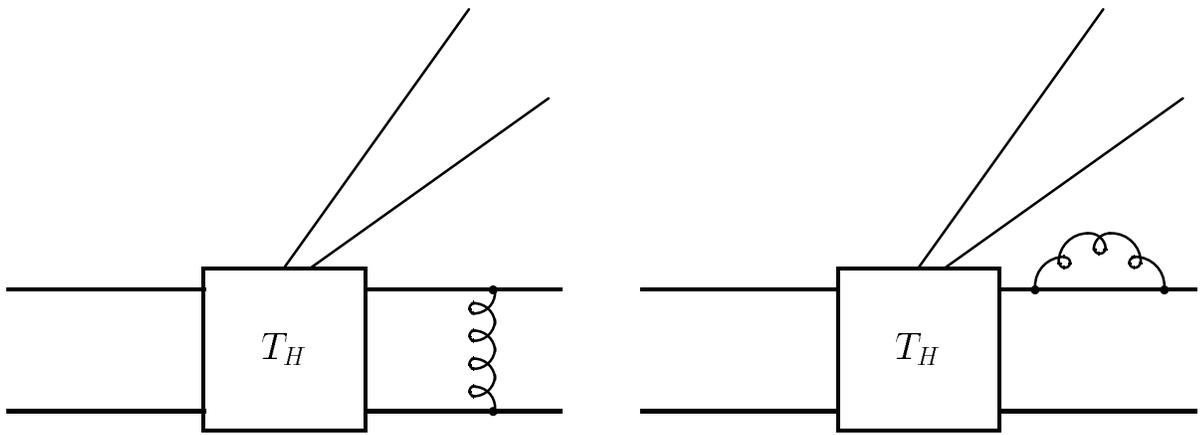
\end{document}